%% file: SAC_paper.tex
\documentclass[preprint,3p,times,twocolumn]{elsarticle}

\input{preamble}

\journal{Nuclear Instruments and Methods in Physics Research A}

\begin{document}

\begin{frontmatter}


\title{Characterization and Performance of PADME's Cherenkov-Based Small-Angle Calorimeter}



\author[cornell]{A. Frankenthal\corref{cor1}}
\author[cornell]{J. Alexander}
\author[lnf]{B. Buonomo}
\author[lnf]{E. Capitolo}
\author[lnf]{C. Capoccia}
\author[cornell,harvard]{C. Cesarotti}
\author[lnf]{R. De Sangro}
\author[lnf]{C. Di Giulio}
\author[infnroma]{F. Ferrarotto}
\author[lnf]{L. Foggetta}
\author[sofia,lnf]{G. Georgiev}
\author[lnf]{P. Gianotti}
\author[hungary]{M. Hunyadi}
\author[sofia,lnf]{V. Kozhuharov}
\author[hungary]{A. Krasznahorkay}
\author[infnroma]{E. Leonardi}
\author[sapienza,infnroma]{G. Organtini}
\author[lnf]{G. Piperno}
\author[sapienza,infnroma]{M. Raggi}
\author[infnroma]{C. Rella}
\author[lnf]{A. Saputi}
\author[lnf,guglielmo]{I. Sarra}
\author[lnf]{E. Spiriti}
\author[lnf]{C. Taruggi}
\author[infnroma]{P. Valente}

\address[cornell]{Cornell University, Ithaca, NY 14853, USA}
\address[harvard]{Harvard University, Cambridge, MA 02138, USA}
\address[sapienza]{Sapienza Università di Roma, Rome, RM 00185, Italy}
\address[infnroma]{INFN Sezione di Roma, Rome, RM 00185, Italy}
\address[lnf]{INFN Laboratori Nazionali di Frascati, Frascati, RM 00044, Italy}
\address[sofia]{University of Sofia, Sofia, 1504, Bulgaria}
\address[hungary]{Institute for Nuclear Research, Hungarian Academy of Sciences, Debrecen, H-4026, Hungary}
\address[guglielmo]{Università degli Studi Guglielmo Marconi, Roma, RM 00193, Italy}

\cortext[cor1]{Corresponding author. \\
Email address: as2872@cornell.edu}

\begin{abstract}
The PADME experiment, at the Laboratori Nazionali di Frascati (LNF), in Italy, will search for invisible decays of the hypothetical dark photon via the process $e^+e^-\rightarrow \gamma A'$, where the $A'$ escapes detection. The dark photon mass range sensitivity in a first phase will be 1 to 24 MeV. We report here on performance measurements and simulation studies of a prototype of the Small-Angle Calorimeter, a component of PADME's detector dedicated to rejecting 2- and 3-gamma backgrounds. The crucial requirement is a timing resolution of less than 200 ps, which is satisfied by the choice of PbF$_2$ crystals and the newly released Hamamatsu R13478UV photomultiplier tubes (PMTs). We find a timing resolution of 81 ps (with double-peak separation resolution of 1.8 ns) and a single-crystal energy resolution of 10\% at 550 MeV with light yield of 2.05 photo-electrons per MeV, using 100 to 400 MeV electrons at the Beam Test Facility of LNF. We also propose the investigation of a two-PMT solution coupled to a single PbF$_2$ crystal for higher-energy applications, which has potentially attractive features.

\end{abstract}


\end{frontmatter}


\section{Introduction}
\label{sec:intro}

Despite the recent discovery of the Higgs boson at the LHC, solidifying our understanding of the standard model (SM) of particle physics, many lingering questions remain that the SM cannot yet explain. One of the prime examples is the dark matter problem -- the fact that we have abundant evidence for its existence yet few clues about its composition.

One proposed solution to the dark matter problem invokes the existence of a dark sector, coupled to the SM sector by its
`dark photon', a vector mediator of a new abelian gauge force that kinetically mixes with the SM $U(1)_Y$. The kinetic mixing coupling can be small, allowing the dark photon to evade discovery so far.

\begin{figure*}[htb]
  \centering\includegraphics[width=\linewidth]{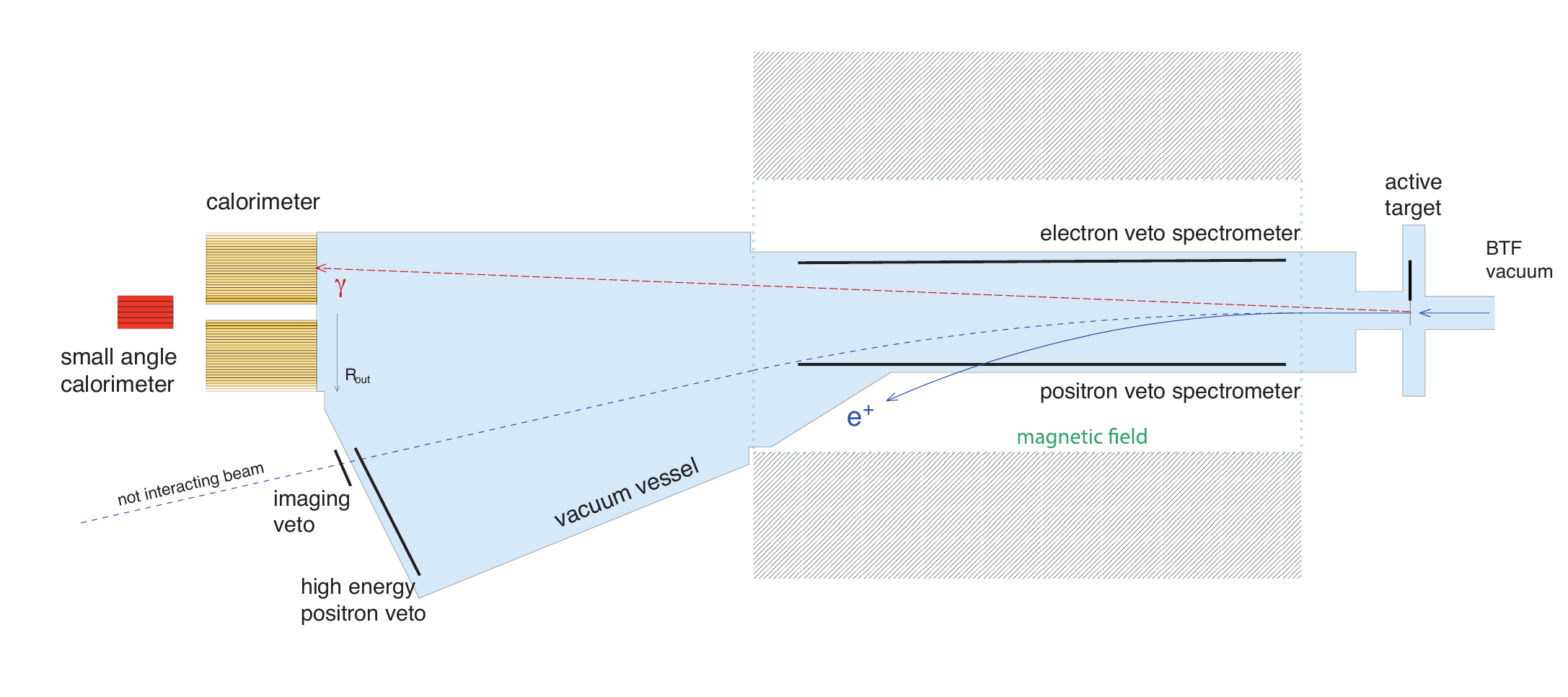}
  \caption{Layout of the PADME detector. On the left is the Small-Angle Calorimeter (in red), which sits behind the central hole in the main calorimeter. Other components include an electron and positron veto spectrometer, an active diamond target, and a \SI{0.5}{\tesla} magnet.}
  \label{fig:padme_general_setup}
\end{figure*}

The Positron Annihilation into Dark Matter Experiment (PADME), set to start operation at the INFN Laboratori Nazionali di Frascati (LNF), in Italy, during the summer of 2018, is a novel fixed-target, missing-mass experiment to search for the dark photon via its effective coupling to ordinary SM electromagnetism  \cite{Raggi:2014zpa}. The goal is to look for remnants of the interaction $e^+ e^- \rightarrow \gamma A'$, where $A'$ is the dark photon.

A beam of \SI{550}{\MeV} positrons is produced and accelerated by a LINAC at LNF's Double Annular $\Phi$ Factory for Nice Experiments (DA$\Phi$NE), and driven to one of two Beam Test Facility (BTF) \cite{Valente:2016tom} magnetic lines. The beam strikes a low-Z target, and the interaction with atomic electrons produces a photon and a dark photon. The energy and direction of the outgoing photon are measured with an electromagnetic calorimeter (ECAL). Combined with an accurate knowledge of the positron beam, the missing mass of the system (i.e. the dark photon invariant mass) can be inferred.

The ECAL has a central hole due to the high rate of Bremsstrahlung events, which are sharply peaked at small angles. In order to have a manageable rate in the ECAL and in the scintillating bars acting as veto detectors for Bremsstrahlung events, the intensity of the positron beam must be limited to \num{2.5e4} $e^+$ in \SI{200}{\ns} long LINAC pulses. 

The proposed experiment requires excellent background rejection capability. Among the largest sources of background are 2-gamma ($e^+e^-\rightarrow \gamma\gamma$) and 3-gamma ($e^+e^-\rightarrow \gamma\gamma\gamma$) events, where 1 (or 2) photons escape detection via the hole in the ECAL. To mitigate such backgrounds, a very fast Small-Angle Calorimeter (SAC) is placed behind the main ECAL, flush with the central hole. The in-time correlation of photon events in the SAC and ECAL allows the tagging of 2- and 3-gamma events and hence the efficient vetoing of such backgrounds.

In this paper we evaluate the performance of a prototype of the SAC with a test beam done at LNF, using fast Lead Fluoride crystals (\pbfII) and the newly developed Hamamatsu R13478UV photomultiplier tube (PMT), optimized for fast response. We demonstrate that this detector meets the requirements for an efficient rejection of 2- and 3-gamma events, namely: (a) a timing resolution less than \SI{200}{\ps} for Cherenkov radiation detection; (b) moderate single-crystal energy resolution better than 10\% at close to beam energy; (c) moderate light yield between 0.5 and 2 photo-electrons (p.e.)/MeV; (d) double-peak separation resolution capable of distinguishing several dozen photons in a \SI{200}{\ns} time span; (e) radiation hardness of order 1 Gy per \num{e13} positrons on target; and (f) acceptance of low-wavelength light due to the Cherenkov spectrum \cite{ferrarotto2018padme}. Furthermore, we encourage the investigation of a related (but more expensive) setup which uses two compact ultra-fast PMTs (R9880U-110) coupled to a single \pbfII{} crystal in order to provide independent efficiency measurements and timing references, for higher-energy applications.

\section{The PADME small-angle calorimeter}
\label{sec:sac}

\cref{fig:padme_general_setup} shows the general layout of the PADME detector, including SAC placement behind the main calorimeter. The SAC consists of 25 \pbfII{} crystals, each with transverse dimensions \SI{30 x 30}{\mm}, and length \SI{140}{\mm}. The total transverse area is therefore \SI{150 x 150}{\mm}, slightly larger than the central square hole of the ECAL. The non-interacting beam is diverted to an off-axis beam dump by means of a \SI{0.5}{\tesla} magnet. The photon rate in the central crystal due to Bremsstrahlung is expected to reach several hundred MHz, depending on beam intensity.

The lateral surfaces of each crystal are wrapped with \SI{50}{\um} thick black Tedlar to minimize optical cross-talk and the back surfaces are coupled to Hamamatsu R13478UV PMTs via UV transparent optical grease, with matching index of refraction for optimal light transmission. We describe below the investigation that led to the choices of crystal and PMT, as it offers some interesting directions for future similar experiments.

\section{Crystal/glass and PMT choices}
\label{sec:choices}

\begin{figure}[hbt]
  \centering\includegraphics[width=\linewidth]{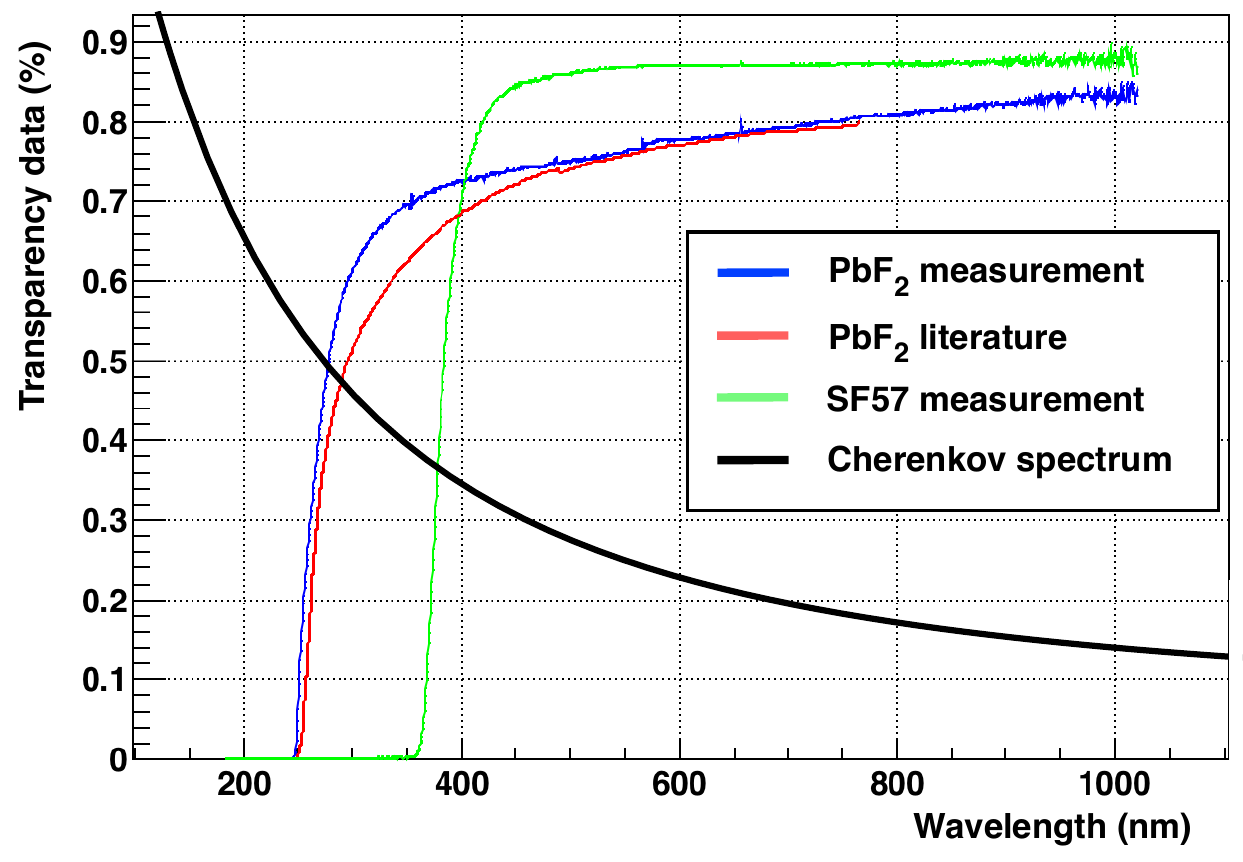}
  \caption{\pbfII{} (red and blue) and SF57 (green) transparency profiles, compared to the Cherenkov spectrum (black, not to scale). For \pbfII{}, our own measurement (blue) agrees well with the literature (red) \cite{fienberg2015studies}.}
  \label{fig:sf57_vs_pbf2_transp}
\end{figure}

Given the requirements for SAC performance outlined in \cref{sec:intro}, two options of radiating material were considered: SF57 (used e.g. in the Large Angle Veto of the NA62 experiment at CERN 
\cite{NA62:2017rwk}) and \pbfII{} (used e.g. in the calorimeters of the Muon g-2 experiment at Fermilab \cite{Grange:2015fou,Anastasi:2016luh} and in the segmented calorimeter of the A4 experiment at MAMI \cite{baunack2011real}). Both materials are suitable Cherenkov radiators due to their high refractive index, allowing for good timing resolution. However, SF57 has two (related) main disadvantages compared to \pbfII. The SF57 transparency window cuts off at \SI{450}{\nm} 
wavelength, whereas the Cherenkov spectrum is peaked at lower wavelengths due to its $1/E_\gamma$ energy dependence. This can be seen in \cref{fig:sf57_vs_pbf2_transp}, which shows the measured transparency profiles (measured by the Atomki Lab in Debrecen) for SF57 and \pbfII{} compared with the Cherenkov spectrum. For this reason, SF57 offers a low light yield: in our preliminary tests, we obtained \SI{0.15}{\pe/\MeV}, which disfavors its use at energies below \SI{1}{\GeV}.\footnote{However, note that this measurement was also limited by the PMT's small dimension (model R9880U-110, described below).}

Furthermore, \pbfII{} is denser and hence more compact, and 10x more radiation-hard.  \cref{tab:sf57_pbf2_comp} contrasts some \pbfII{} and SF57 properties which are relevant for Cherenkov-based calorimetry. \pbfII{} has a shorter radiation length and smaller Moliere radius compared to SF57. Smaller electromagnetic showers reduce the detector occupancy, thus enhancing its rate capabilities.
Hence, \pbfII{} is the preferred solution for PADME's requirements.
	
\begin{table}[htb]
\caption{Comparison between \pbfII{} and SF57 of some properties relevant to Cherenkov calorimetry.}
\label{tab:sf57_pbf2_comp}
\begin{center}
\centering
\begin{tabular}{ c | c | c | c }
\hline
\hline
Property & \pbfII{} & SF57 & Units \\
\hline
Density & 7.77 & 5.51 & g/cm$^3$ \\
$X_0$ & 0.93 & 1.54 & cm \\
Moliére Radius & 2.12 & 2.61 & cm \\
Interaction Length & 22.1 & 20.6 & cm \\
$\lambda/X_0$ & 23.65 & 13.3 & N/A \\
n & 1.8 & 1.8 & N/A \\
\hline
\hline
\end{tabular}
\end{center}
\end{table}

\begin{figure*}[htb]
  \centering\includegraphics[width=\linewidth]{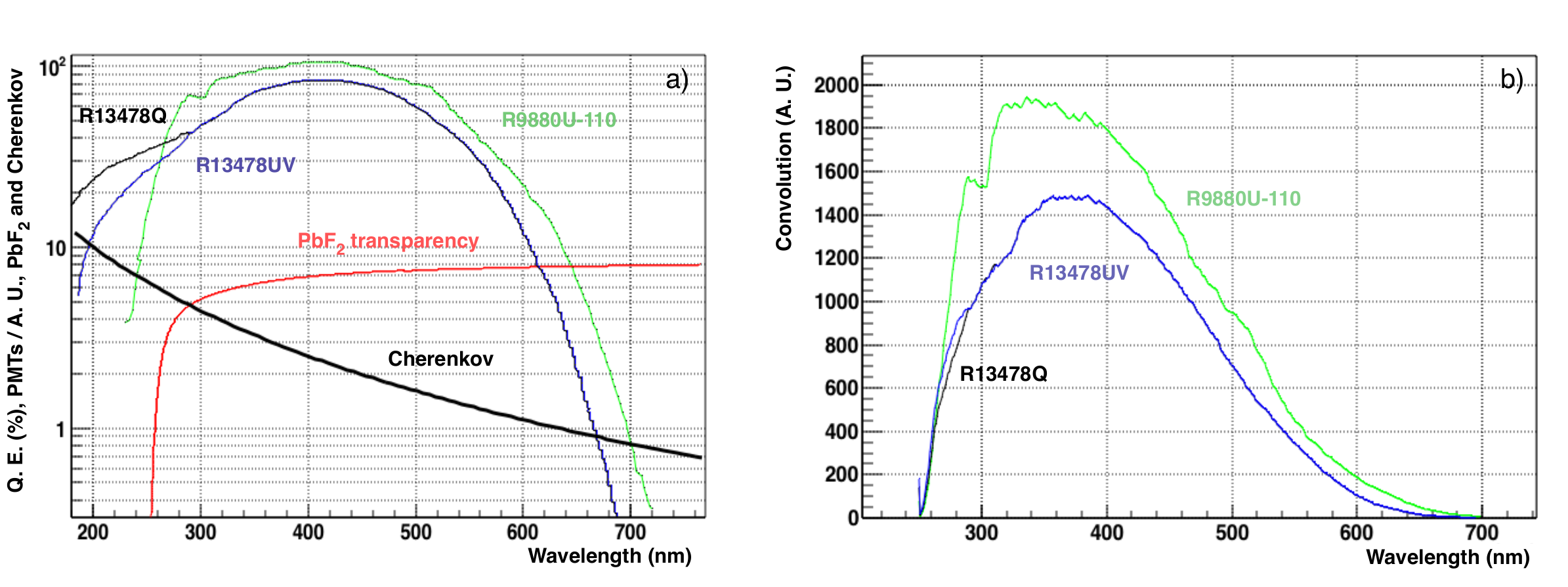}
  \caption{(a) Quantum efficiency (QE) curves for three different PMTs: R13478Q (black), R13478UV (purple), and R9880U-110 (green) \cite{hpkr13478, hpkr9880u}. The Cherenkov spectrum (bold black) and \pbfII{} transparency (red, \cite{fienberg2015studies}) are also plotted for comparison. (b) Convolution of each QE curve with \pbfII{} transparency and Cherenkov spectrum. Due to the drop in transparency below \SI{250}{\nm} there is effectively no difference between the two R13478 PMT models. The R9880U-110 PMT has about 30\% better performance over the entire spectrum.}
  \label{fig:fom_and_conv}
\end{figure*}

We also considered two candidates for PMT in the SAC: the ultra-compact Hamamatsu R9880U-110, and the Hamamatsu R13478 (both Q and UV versions). The first option is a compact PMT with only \SI{16}{\mm} in diameter (\SI{8}{\mm} sensitive area), and a rise time of \SI{0.57}{\ns} and a transit time of \SI{0.2}{\ns}. The small size of this PMT allows the coupling of two such devices to the back of a single crystal. This could provide several benefits, such as improved light yield and an independent time reference between the two PMTs. Furthermore, having two PMTs enables an efficiency measurement of each one separately.

The disadvantage of the R9880U-110 PMT, however, is that its small dimensions limit single-PMT light collection efficiency. Without employing the two-PMT solution, a single R9880U-110 PMT has a cross-sectional area of only 5.5\% the crystal transverse dimensions. Even coupling two of them provides only 11\% of geometric acceptance, while doubling the cost. Given PADME's relatively low beam energy, this light yield is unacceptably small. Nevertheless, for higher energy applications, the two-PMT setup could be an interesting solution to explore.

The second candidate considered was the Hamamatsu R13478, Q and UV versions. This PMT has a diameter of \SI{26}{\mm} (\SI{17.3}{\mm} sensitive area) and thus covers a larger area fraction overall (26\%). Compared to the R98880U-110, it has a similar rise time of \SI{0.9}{\ns} but a slower transit time of \SI{9.1}{\ns}. These specifications are fast enough for the PADME use case, and the improved light collection efficiency enables only one PMT per crystal.

\cref{fig:fom_and_conv} illustrates some key measures of performance underscoring the determination of the optimal PMT. On the left are the quantum efficiency (QE) curves for the different PMTs, as well as the \pbfII{} transparency profile and Cherenkov spectrum. The plot on the right shows the convolution of \pbfII{}, Cherenkov, and QE curves for each PMT. Due to the drop in \pbfII's transparency around a wavelength of \SI{250}{\nm}, there is no significant difference in performance between the more expensive R13478Q model and the more affordable R13478UV. The compact R9880U-110 actually performs better than the R13478's by a factor of roughly 30\%, but its reduced light acceptance must also be accounted for as mentioned above.

To decide between the R9880U-110 and the R13478UV PMTs, we completed a study to understand the signal distribution expected for each one. We considered two configurations: the R13478UV PMT with a tapered voltage divider, with typical gain $G$ of \num{3.2e5}, and the more compact R9880U-110, with a typical gain of \num{2e6}. Both PMTs are assumed to be coupled to a crystal producing $\sim 20$ Cherenkov photons reaching the photosensors face ($N_{ph}$) for each MeV of deposited energy $E_{dep}$, and to have average 20\% QE. To account for differences between the two PMTs, two corrections are added depending on the PMT. The light acceptance due to photocathode size $A_C$ corrects for the different active areas, while $QE_{corr}$ corrects for the different integrated QEs (see \cref{fig:fom_and_conv}). First we calculate the number of p.e. produced by each configuration:

\begin{equation} N_{p.e.} = N_{ph}\times E_{dep} \times QE \times A_C \times QE_{corr} .
\end{equation}

With these assumptions, we estimate the expected charge distribution 
from an incident particle based on a Gaussian spread model:
\begin{equation} Q_{tot} = \textrm{Gaus}(N_{p.e.}, \sqrt{N_{p.e.}})\times G\times e .
\end{equation}

\begin{figure}[hbt]
  \centering\includegraphics[width=\linewidth]{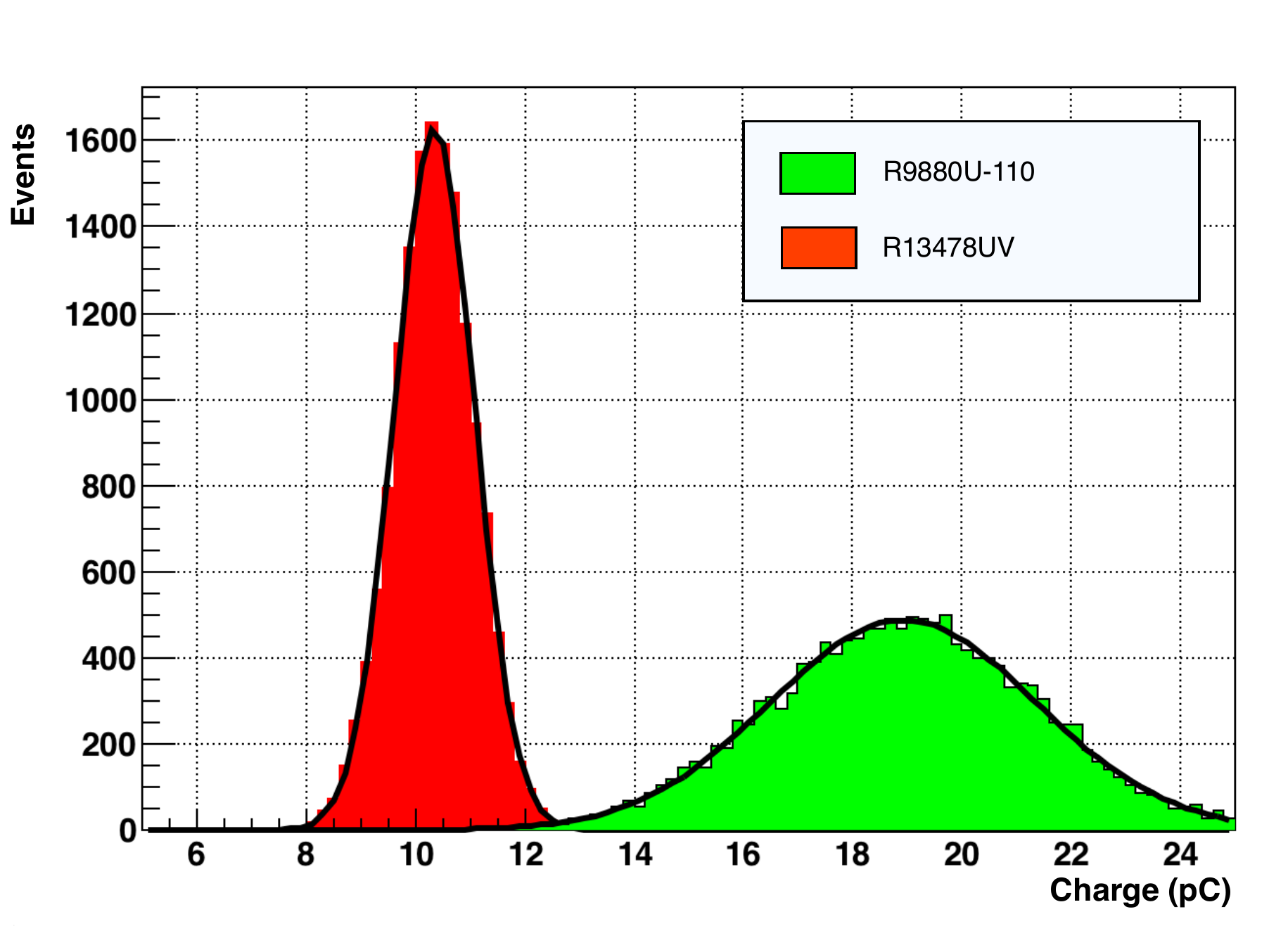}
\caption{Simulation of the charge distribution expected for each PMT for a given deposited energy. R13478UV (red) and R9880U-110 (green) PMTs are compared.}
\label{fig:charge_dist}
\end{figure}

The results are shown in \cref{fig:charge_dist}, with the charge distributions and corresponding estimated resolutions. Despite a higher gain, the R9880-110's small surface area limits light collection efficiency and hence the charge resolution. The R13478UV PMT offers in this configuration roughly a factor of 2 better resolution.

From these considerations, we conclude the larger Hamamatsu R13478UV PMT, coupled to \pbfII{} crystals, is the best solution for PADME's requirements. This has acceptable timing resolution while providing sufficient light yield for the low beam energies in the experiment, at a reasonable cost. 

\section{Monte Carlo simulation and radiation damage}

We developed a full Monte Carlo (MC) simulation of the PADME detector with the Geant4 framework \cite{agostinelli2003geant4} to study different aspects of the experiment. Three major goals discussed in this paper include assessing the expected levels of absorbed radiation dose in the calorimeter, studying the detector response as a function of crystal properties, and correcting for the average shower energy leakage in our single-crystal test beam. In the first case, we simulate the entire physics, starting from beam positrons. In the second, we directly simulate high-energy photons incident on the crystals. And in the third, we simulate electrons striking a single crystal in order to match accurately the electron test beam results. In this section we describe our radiation damage study, while the other two questions are addressed further below.

The full MC simulation contains all relevant components of the detector, such as target, magnet, ECAL and SAC, and veto spectrometer. For the radiation dose study, we generated 400 million events consisting of positrons with energy \SI{550}{\MeV} striking the target. The resulting photons (mostly from positron Bremsstrahlung emission, but also from 2- and 3-gamma production) then strike the calorimeters. The spectrum of Bremsstrahlung radiation is highly peaked at small angles, so the SAC receives most of the radiation. We then re-scaled the statistics of simulated data up to the expected integrated luminosity of $1\times 10^{13}$ positrons on target over the course of the experiment. 

\begin{figure}
    \centering
    \includegraphics[width=\linewidth]{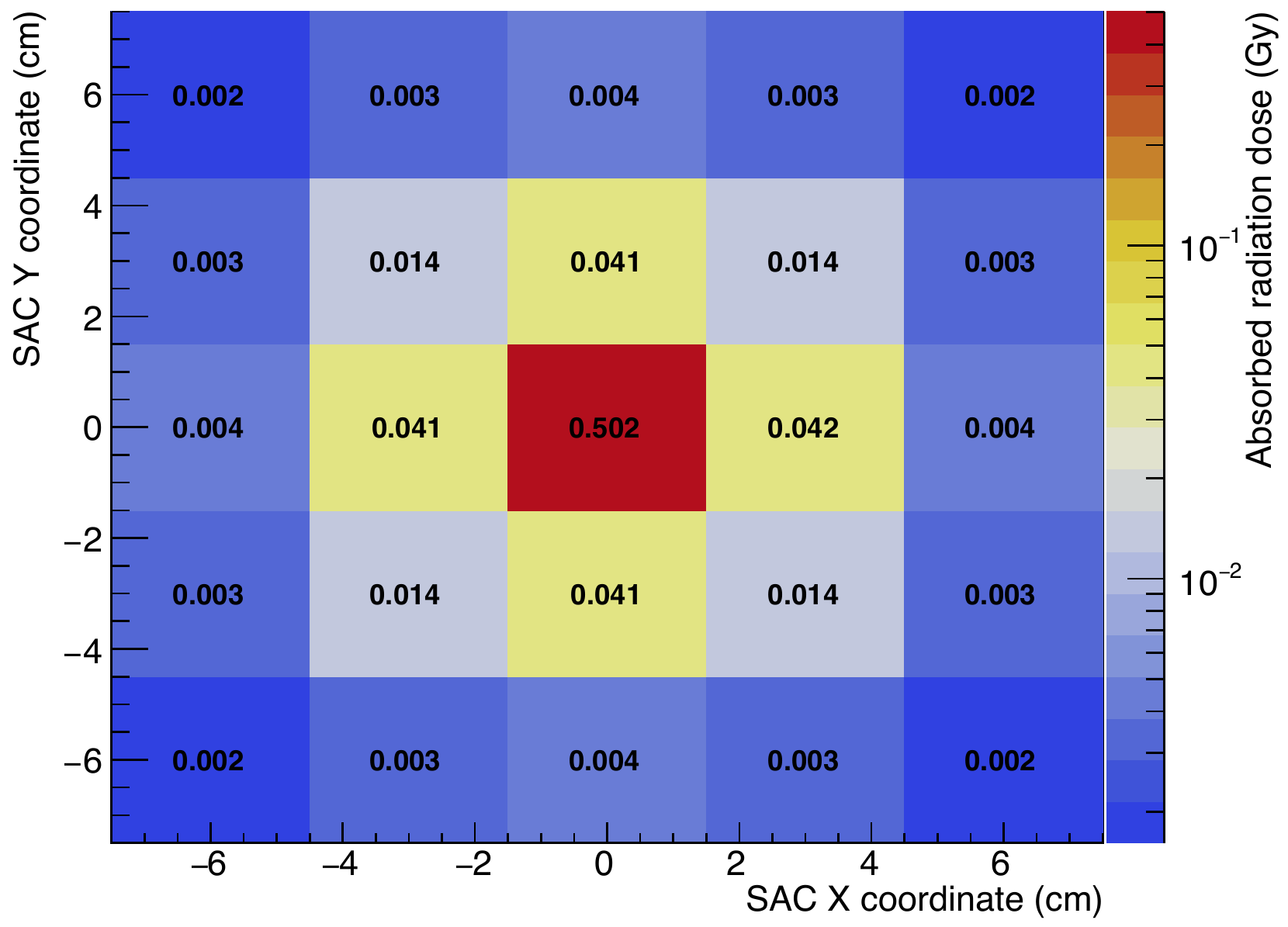}
    \caption{Expected radiation dose on SAC crystals from $10^{13}$ positrons on target, over the course of the experiment, as estimated by a Geant4 MC simulation. The center-most crystal (which receives the most radiation due to Bremsstrahlung's sharply peaked angular spectrum) should absorb a dose of only $\sim$ 1 Gy.}
    \label{fig:rad_dose}
\end{figure}

\cref{fig:rad_dose} shows the total expected radiation dose absorbed by SAC crystals. This estimate is made by assuming that the energy of striking photons is entirely transferred to the crystals, and that the energy deposition is uniform across the crystal. The dose is then obtained by dividing the total energy deposit by the mass of each crystal. The assumption of uniformity is not accurate, as most of the energy deposits happen early on during the shower development inside a crystal \cite{wigmans2002differences}. Nevertheless, we are only interested in an order-of-magnitude estimate of radiation dose, to show that radiation damage on the crystals will be negligible throughout the experiment and that we can safely mitigate concerns about transparency losses.

\begin{figure}[hb]
    \centering
    \includegraphics[width=\linewidth]{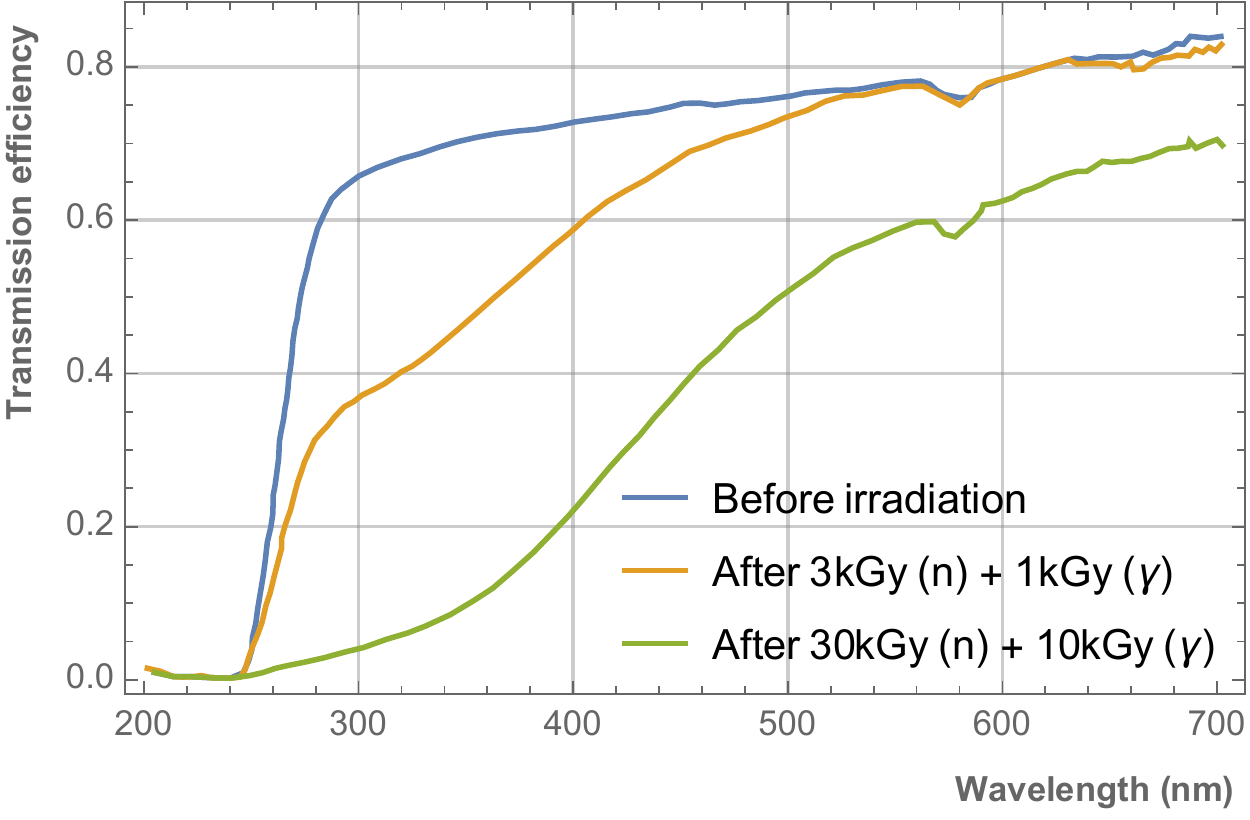}
    \caption{Transmission efficiency data before and after \SI{4}{kGy} and \SI{40}{kGy} of irradiation with neutrons and gamma rays \cite{anderson1990lead}, for a \SI{1}{cm^3} \pbfII{} cube. Comparing with our expected radiation doses (\cref{fig:rad_dose}), the transmission loss for PADME crystals should be negligible. }
    \label{fig:transp_loss}
\end{figure}

There are several studies of transmission loss in \pbfII{} crystals due to radiation damage in the literature, e.g. \cite{barysevich2013radiation} (protons) and \cite{anderson1990lead} (neutrons and gamma rays). We show the latter's \pbfII{} transparency data in \cref{fig:transp_loss}, where transmission efficiency as a function of wavelength is plotted before irradiation, and after 4 kGy and 40 kGy of combined irradiation with neutrons and gamma rays. Comparing this data with our SAC crystal doses, we therefore predict negligible transparency loss after the full run period of the PADME data-taking.

\section{Optical MC simulation}
\label{sec:optical_mc}

To establish our crystal specifications and to study different aspects of light collection efficiency, timing and energy resolution, we adapted our detector MC simulation to include a detailed optical description of physics inside the crystals. The primary goal was to investigate differences in light yield and photon arrival times as a function of crystal length and determine the optimal length. Manufacturing limitations restrict the crystal's length to 18 cm.

This simulation consists of a \pbfII{} crystal, with transverse dimensions \SI{30x30}{\mm}, and a variable length in the range \SIrange{10}{20}{\cm}. It can be wrapped with either white millipore or black tedlar materials. The crystal is coupled to a thin layer of grease, modeled as Epoxy EJ-500 (radius \SI{12.7}{\mm}, thickness \SI{1}{\mm}), and then to a sensitive detector layer which models the PMT. The QE of the R13478UV PMT is implemented as a function of energy according to its datasheet \cite{hpkr13478}. Other optical properties are also implemented as functions of energy, such as absorption length from \pbfII{} transparency \cite{fienberg2015studies}, and refractive index via a parametrized dispersion formula \cite{malitson1969refraction}. These were calculated for optical photons with energy ranging from \SIrange{1.6}{5.0}{\eV} in steps of \SI{0.02}{\eV}. Note that we do not simulate the entire detector here, just a single crystal in the SAC.

A single energetic photon (energy: \SI{200}{\MeV}) is fired at a distance of \SI{1}{\mm} from the crystal's front surface, which produces a few thousand optical photons after showering inside the crystal. The simulation then tracks each individual optical photon, until it either reaches the sensitive area corresponding to the PMT, or gets lost along the way. We run 100 events per crystal length.

\begin{figure}[hbt]
  \centering\includegraphics[width=\linewidth]{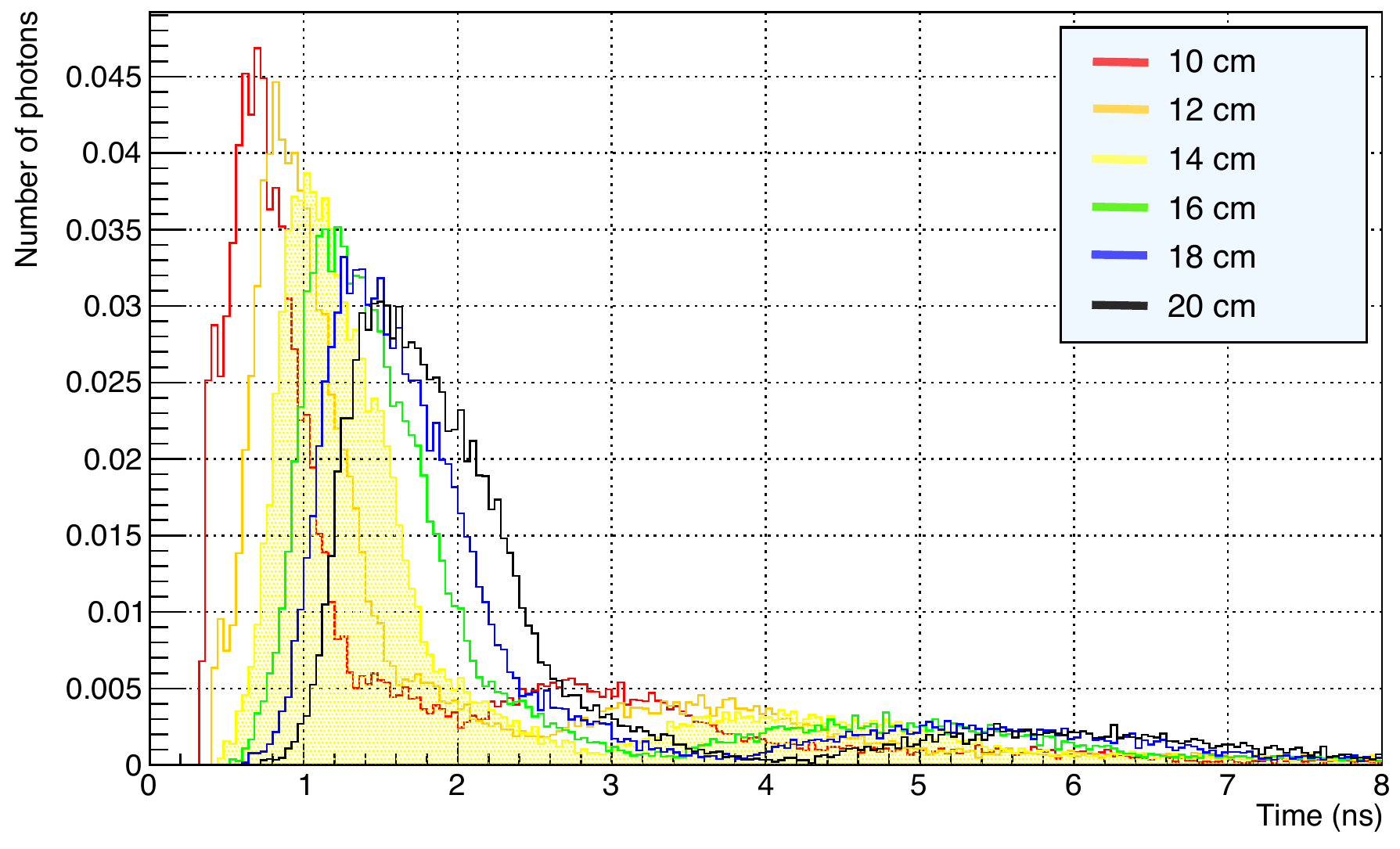}
  \caption{Arrival time distribution of Cherenkov photons versus crystal length for a \SI{200}{\MeV} incident photon. The distribution shifts to the right and broadens with increasing length.}
  \label{fig:arr_time_vs_length}
\end{figure}

The arrival time distribution of photons for each length considered is displayed in \cref{fig:arr_time_vs_length}. Two peaks can be identified in each curve: a narrow, high peak which represents the arrivals of most photons, and a lower, broader peak, which corresponds to photons that underwent back-scattering inside the crystal. A shift and broadening of the narrow peak with increasing length is clearly visible.

\begin{figure}[hbt]
  \centering\includegraphics[width=\linewidth]{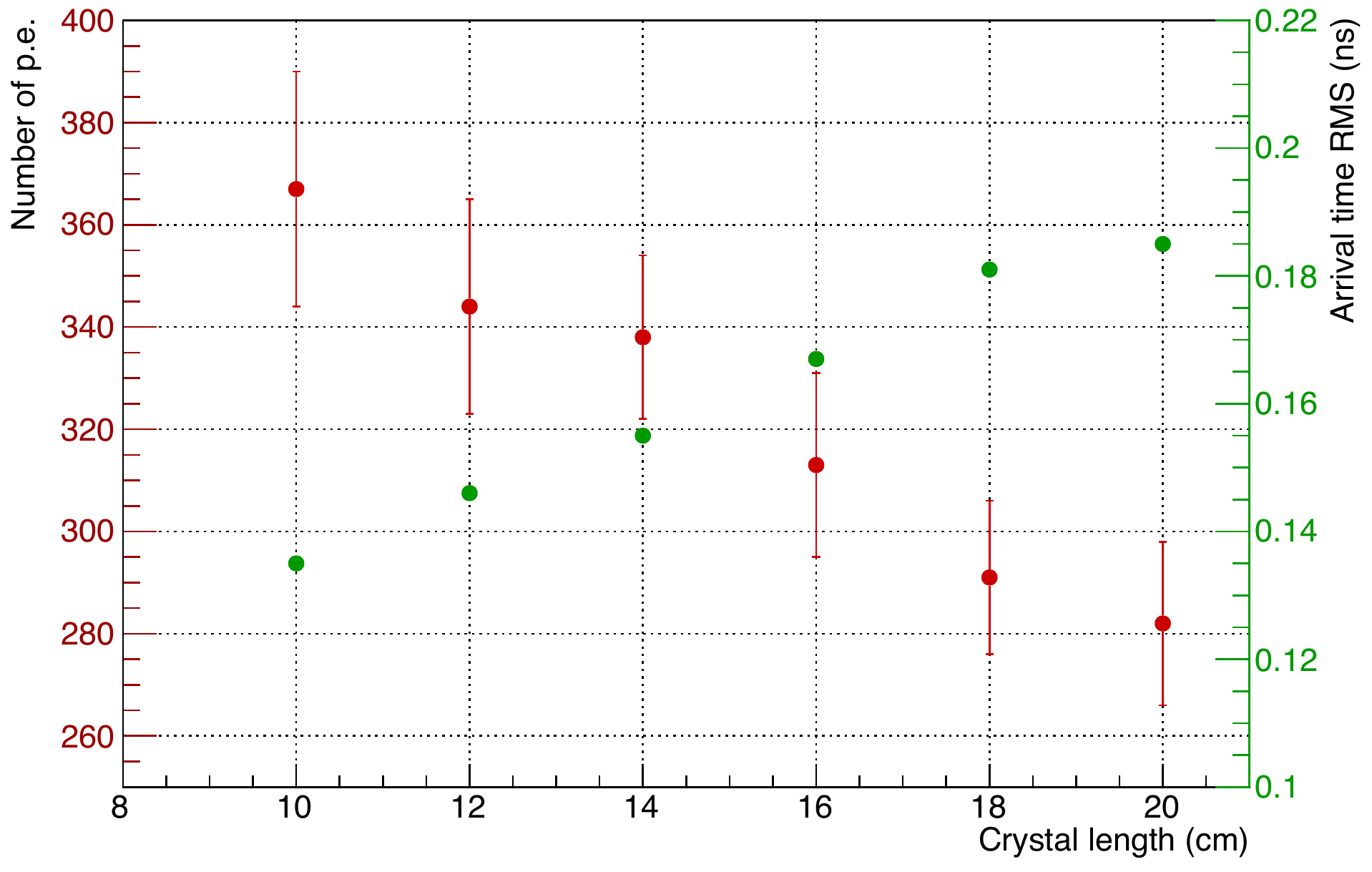}
  \caption{Light yield (red) and Landau time spread (green) versus crystal length for a \SI{200}{\MeV} incident photon, based on a detailed optical MC simulation. The light yield is estimated by convolving the energy distribution of arriving photons with the PMT's QE profile, and the timing resolution by a Landau fit of the arrival time distribution.}
  \label{fig:ly_and_time_rms}
\end{figure}

The arrival profiles were fitted to a convolution of a Landau (narrow peak) and a Gaussian (broader peak) distribution, and the sigma of the Landau was taken as indicative of the double-particle separation capabilities of the detector in MC. The choice of a Landau fit was empirically driven and not based on underlying physical processes. We do not account for effects from the PMT itself, but those should be of second order and so this procedure at least allows the relative comparison of different crystal lengths. The MC Landau spread time is plotted in \cref{fig:ly_and_time_rms}. There is roughly a 16\% increase when going from \SI{14}{\cm} to \SI{18}{\cm}.

\cref{fig:ly_and_time_rms} also shows the light yield as a function of crystal length. The light yield is determined from a convolution of the PMT's QE with the energy distribution of arriving photons. The light yield decreases with increasing length due to a higher chance of Cherenkov photons getting lost (absorbed or escaping) while traveling towards the PMT. In particular, there is a drop of about 14\% in collected light between lengths of \SI{18}{\cm} and \SI{14}{\cm}.

From the optical MC simulation, it is evident that shorter crystals are better, both for optimal light yield and for timing considerations. Increasing the length causes more Cherenkov photons to be lost along the way. At the same time, in a longer crystal photons have longer travel paths. Consequently, there is more opportunity for scattering, which broadens the arrival time profile of photons. Thus the simulation reveals that increasing the crystal length leads to broader signals and longer tails, affecting the double-peak separation capability, which is a crucial parameter for PADME SAC performance. 

Nevertheless, there is a length limit where the crystal is too short to develop a complete Cherenkov shower. From \cite{fienberg2015studies}, about 50\% of Cherenkov photons from a \SI{3}{\GeV} shower are produced within the first \SI{60}{\mm} of a \pbfII{} crystal, and about 80\% are produced within the first \SI{100}{\mm}. This makes \SI{140}{\mm} a safe length to ensure we can harness most of the Cherenkov photons in a \SI{500}{\MeV} shower.

\section{Test beam setup}

The chosen crystal and PMT options were subjected to a test beam at LNF's BTF, in order to characterize the SAC response and measure energy and timing resolution, including double-peak separation capability. We obtained \pbfII{} crystals on loan from Brookhaven National Laboratory for this test, with dimensions \SI{30x30x140}{\mm}.

The BTF at LNF is part of the  DA$\Phi$NE accelerator complex. A LINAC provides bunches of \SI{\sim e10}{electrons} or \SI{\sim e9}{positrons} with energy up to \SI{750}{\MeV} or \SI{550}{\MeV}, respectively. The LINAC has to switch between electron and positron modes on a regular interval, in order to top off the main DA$\Phi$NE rings, and in this case the energy is fixed to $E_0 = \SI{510}{\MeV}$ for both charges. There are 50 bunches per second exiting the LINAC, which are shared between the BTF and the main ring under normal operation. In this regime, the BTF gets \SI{38}{bunches/\s} in electron mode, and \SI{18}{bunches/\s} in positron mode. Each mode lasts \SI{180}{\s}, and then a 90-second switch mode is activated. During the switch mode, there are \SI{60}{\s} with no beam at all and then \SI{30}{\s} where BTF gets the entire \SI{49}{bunches/\s} (one per second is used for beam-energy monitoring). The cycle then repeats. 

The BTF setup also allows the tuning of the beam intensity delivered to the experimental hall, from \SI{\sim e10}{particles/bunch}, down to a single particle/bunch, by intercepting the primary beam with a 2 $X_0$ target. The resulting secondary particles can be further filtered to allow tuning of the beam energy from $E_0$ down to few tens of MeV. The beam spot and position can be adjusted by means of quadrupoles, dipoles and correctors in the BTF line, and is monitored in real-time by silicon pixel hybrid detectors (FITPIX \cite{1748-0221-6-01-C01079}) with active area \SI{14 x 14}{\mm} and \SI{55}{\um} pitch.

\begin{figure}[hbt]
  \centering\includegraphics[width=\linewidth]{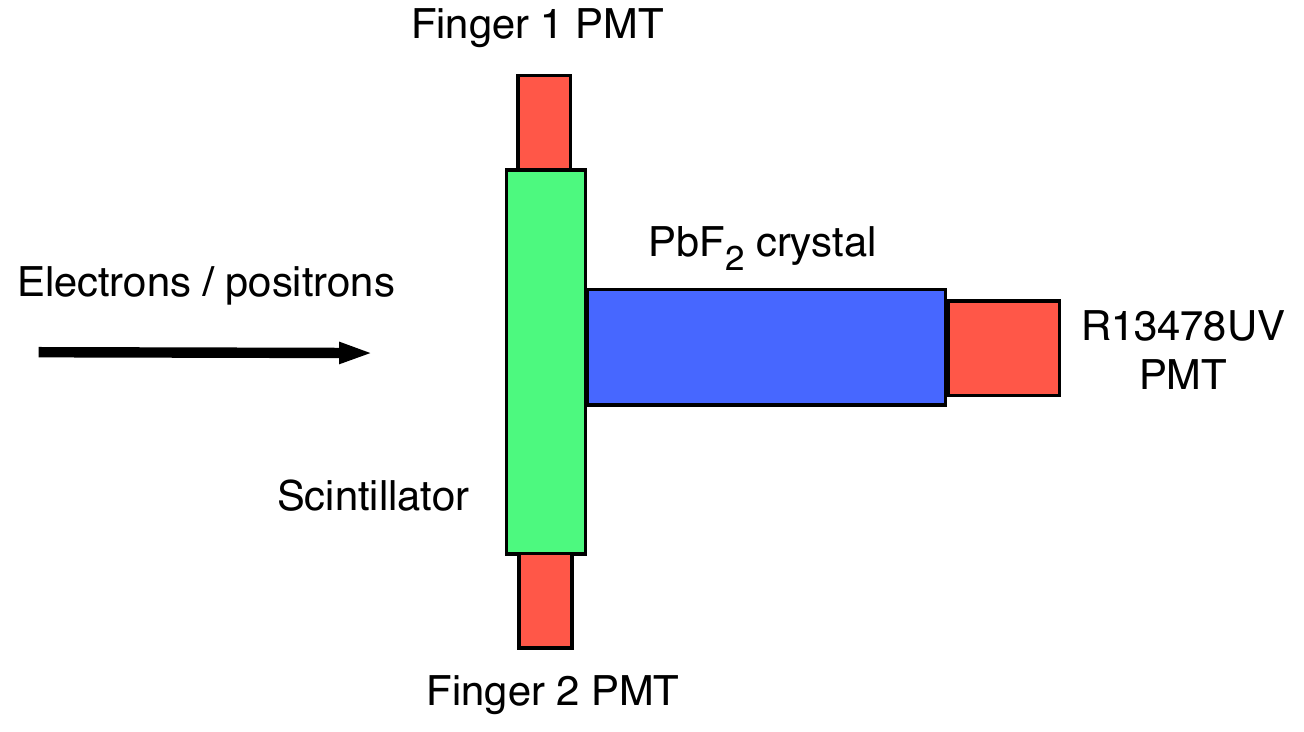}
  \caption{Schematic layout of the test beam detector setup. A \pbfII{} crystal is coupled to the R13478UV PMT via optical grease. Two compact finger PMTs connect to a plastic scintillator bar and provide a coincidence reference signal. The operating voltage of the PMT is \SI{1600}{\V}.}
  \label{fig:tb_setup}
\end{figure}

A schematic of the detector test setup is shown in \cref{fig:tb_setup}. A R13478UV PMT (voltage: \SI{1600}{\V}) was coupled to a \pbfII{} crystal using optical grease and then connected to a 12-bit, \SI{5}{GSPS}, 1024-sample digitizer (model CAEN V1742) for data acquisition. A plastic scintillator coupled to two small `finger' PMTs (model R9880U-110) provided a reference signal for comparison with the \pbfII{} one.

\begin{figure}[hbt]
  \centering\includegraphics[width=\linewidth]{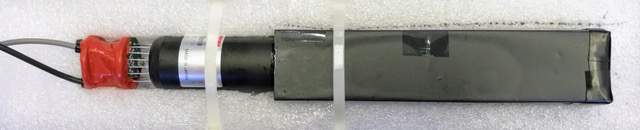}
  \caption{Picture of the PMT and crystal setup for the SAC test beam. A \pbfII{} crystal is coupled to the R13478UV PMT via optical grease.}
  \label{fig:tb_setup_pic}
\end{figure}

A picture of the PMT and crystal setup is shown in \cref{fig:tb_setup_pic}.

For the studies presented here, the beam energy was varied between \SIlist{100;400}{\MeV} in steps of \SI{100}{\MeV}, and the average number of particles per pulse delivered to the BTF experimental hall was set to $\sim$ 1 for electrons. The beam was centered on the crystal and the spot was kept within a standard deviation of 3 mm (at 400 MeV) and 5 mm (at 100 MeV) in the transverse plane.

\subsection{Single-crystal Monte Carlo simulation}
\label{sec:sc_mc}

Since only one \pbfII{} crystal was available for testing, we adapted our dedicated MC simulation to characterize the average energy leakage and provide a correction factor to light yield and energy resolution measurements. 

The simulation in this case consists of a single \pbfII{} crystal with the same dimensions as the one used at BTF and a beam of electrons with similar energy and multiplicity as the test beam. The primary goal was to estimate the average fraction of incident energy that escapes the crystal, decreasing the energy collection efficiency. This is a purely geometrical correction which can be used to re-scale the light yield and energy resolution obtained from test beam data.

\begin{figure}[hbt]
  \centering\includegraphics[width=\linewidth]{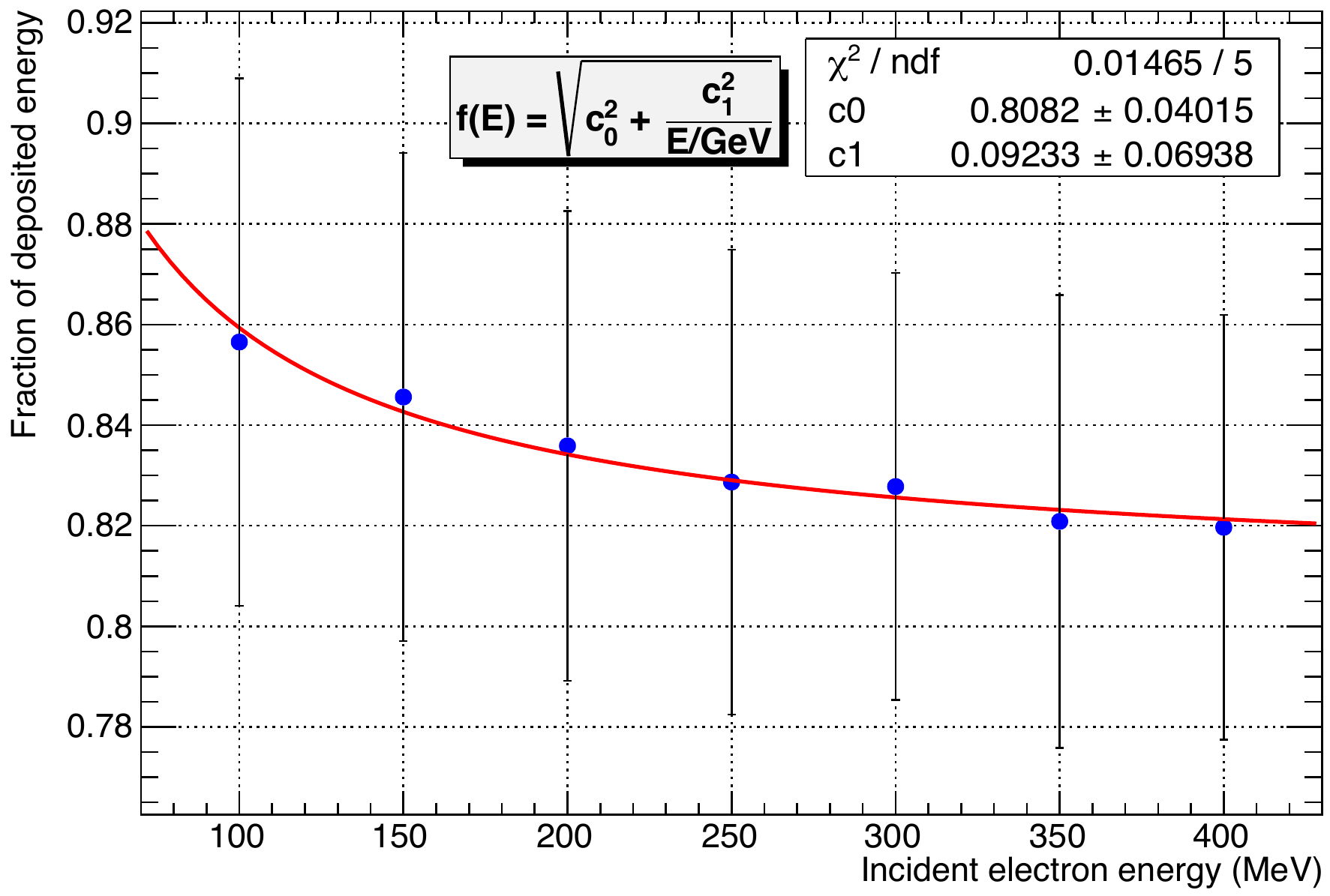}
  \caption{Fraction of deposited energy on a \SI{30x30x140}{\mm} \pbfII{} crystal with a high-energy incident electron, as determined by a Geant4 MC simulation. Each data point is the mean of a Crystal-Ball fit to the distribution of deposited energies with 10k events.}
  \label{fig:energy_deposit}
\end{figure}

\cref{fig:energy_deposit} shows the average fraction of deposited energy in the crystal as a function of incident electron energy. For each energy, 10k events were simulated and the distribution of deposited energy was fit to a Crystal-Ball (CB) function \cite{gaiser1982charmonium}. The extracted mean was used as a data point in \cref{fig:energy_deposit}, and the extracted sigma as the uncertainty in that value.

The plot was fit to the empirical function:

\begin{equation}
    f(E) = \sqrt{c_0^2 + \frac{c_1^2}{E/\text{GeV}}}.
\end{equation}

\noindent where $f(E)$ is the average fraction of energy deposited on the crystal.

Note that this procedure can only correct for average energy leakage, and does not account for experimental fluctuations in that quantity. Therefore the energy resolution quoted here is only an upper limit to the actual one achievable by PADME, and should be interpreted as a single-crystal SAC energy resolution.

\section{Charge reconstruction}

The results presented in this paper are based on data taken in July of 2017 at the LNF BTF. Electron beams with average multiplicity of \SI{1}{particle/pulse} and energies of \SIlist{100;200;300;400}{\MeV} impinged on the detector setup. Data acquisition triggered on accelerator signals, and 1024 samples at \SI{5}{GSPS} (i.e. \SI{0.2}{\ns/sample}) were collected per trigger \cite{drs4development}. Since \pbfII{} has a fast Cherenkov emission of less than a few nanoseconds, the signal is centered within a small window of the waveform.

The total energy deposited in the crystal was reconstructed in three steps. First a run-level pedestal was calculated by averaging the ADC counts of each 1000-sample empty event in the run (i.e. only noise, no signal peaks). The last 24 samples of each event were not used. The average ADC counts for all empty events form a Gaussian distribution, the mean of which is taken as the pedestal for that run. The sigma in turn informs the noise level, which was around \SI{1.1}{\pico\coulomb} for all runs. This is the noise considering all 1000 samples, but for a roughly 50-sample signal window the average noise is scaled down to order \SI{0.1}{\pico\coulomb}.

\begin{figure}[hbt]
  \centering\includegraphics[width=\linewidth]{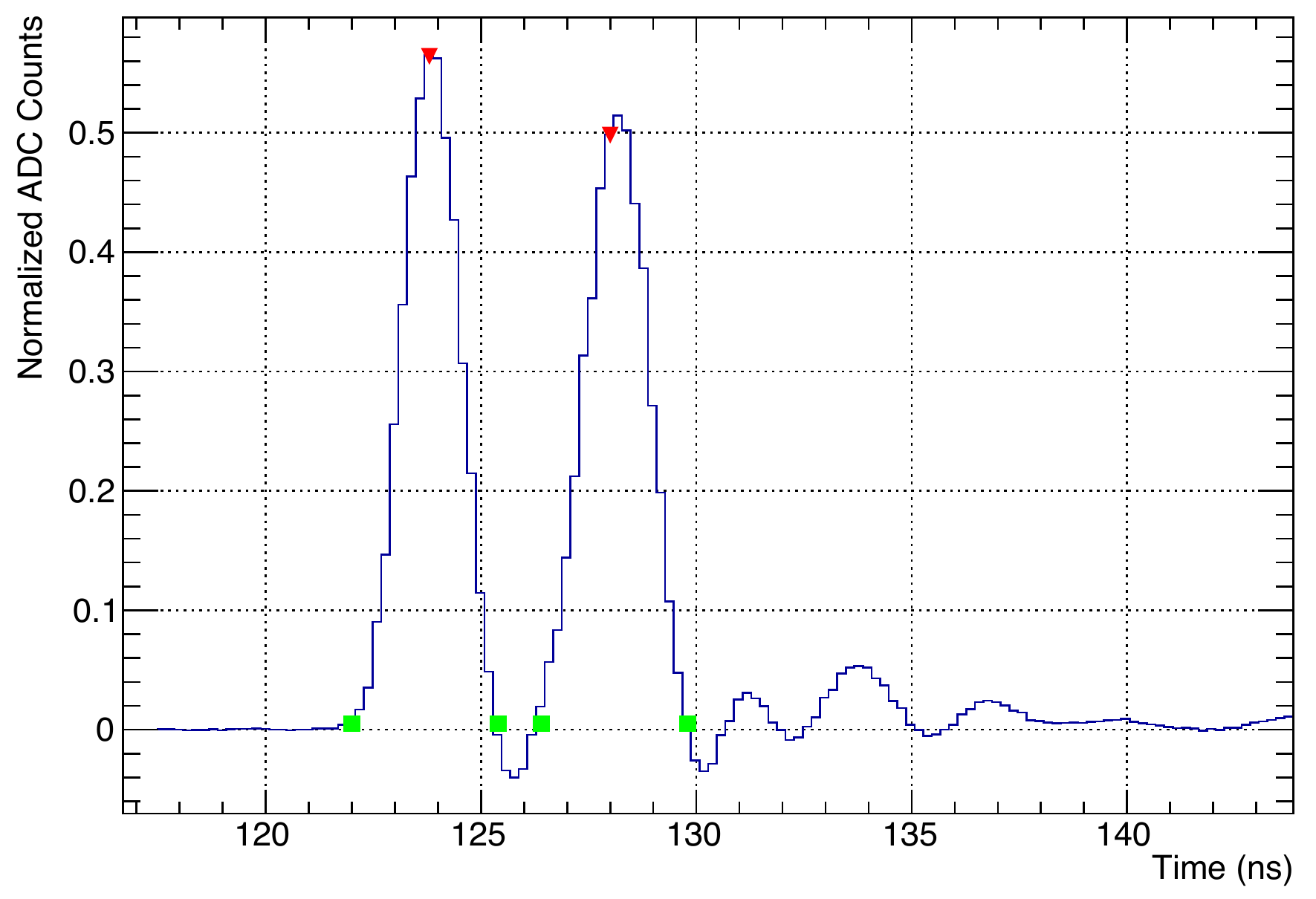}
  \caption{Sample digitized trace with two electron peaks. The peaks are identified with a custom algorithm and confirmed with ROOT's TSpectrum class. For each peak the thresholds are determined by using the after-pulse ringing and setting a low threshold (but still above the average noise level). The time between samples is \SI{0.2}{\ns}.}
  \label{fig:sample_trace}
\end{figure}

After subtracting the pedestal from each ADC count, the integrated charge was calculated by identifying all signal peaks in a given event, and integrating the area underneath each peak. The peak boundaries were set via a simple threshold ($|\mathrm{ADC}-\mathrm{pedestal}|/4096 > 0.005$). A sample digitized trace, with two electron peaks and thresholds identified, is shown in \cref{fig:sample_trace}.

The choice of threshold was made to mitigate the after-pulse ringing which can be seen in the figure. This ringing continues with approximately constant magnitude for about 40 ns after the end of a pulse. Immediately following each pulse, the signal shoots below the pedestal-subtracted zero level, which motivates the low threshold in order to exclude such ringing effects from the charge estimate.

\begin{figure}[hbt]
  \centering\includegraphics[width=\linewidth]{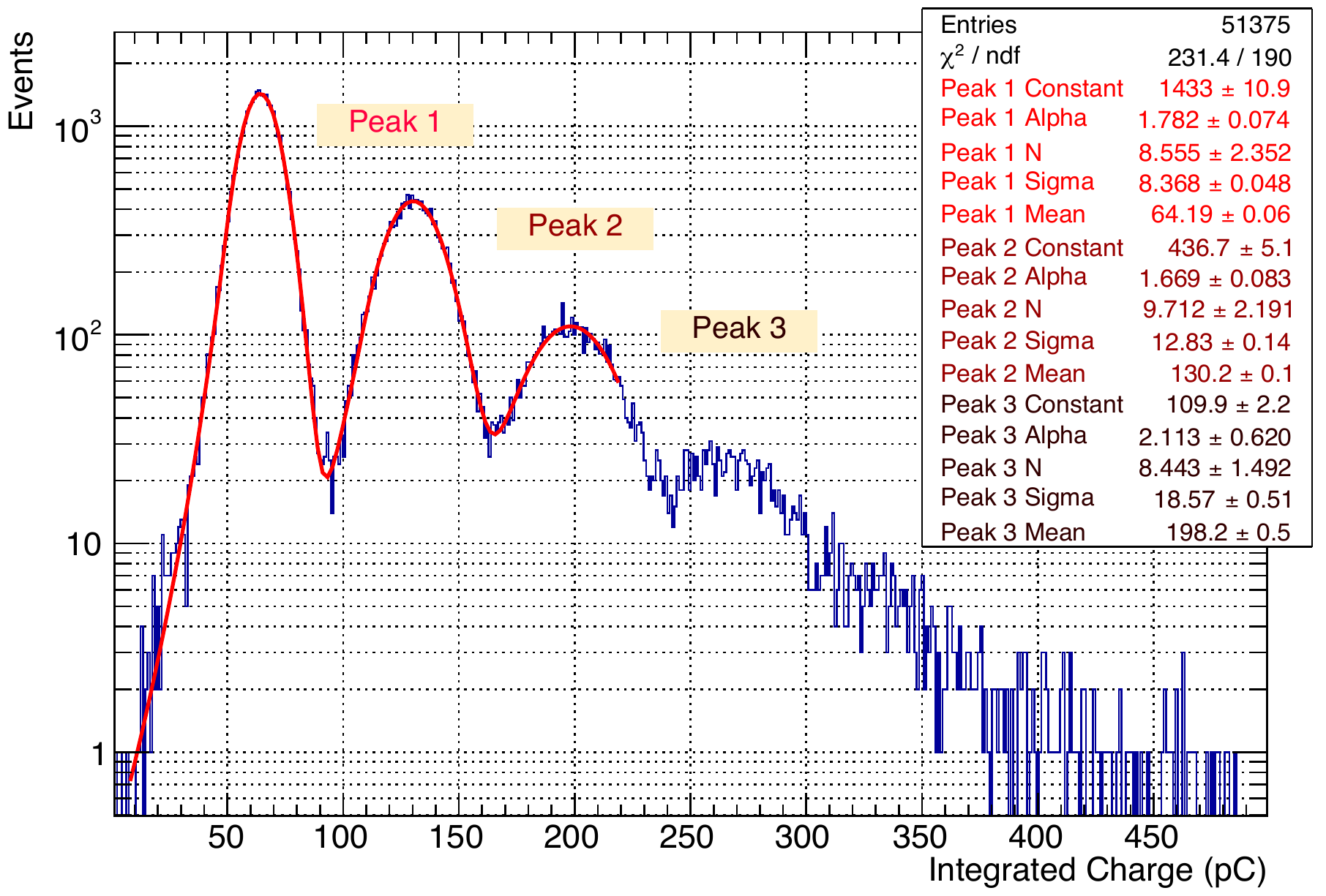}
  \caption{Integrated charge distribution for \SI{300}{\MeV} electrons. The first 3 peaks were fit with a sum of Crystal-Ball functions for extraction of means and sigmas.}
  \label{fig:cb_fits}
\end{figure}

Finally, the distribution of integrated charges was plotted. A representative example, for \SI{300}{\MeV} electrons, is shown in \cref{fig:cb_fits}. The peaks were fit to a sum of CB functions in order to extract the mean and sigma.

\section{Detector performance}

To estimate single-crystal performance, we measured linearity and energy resolution, using only single-electron events, at different incident electron energies. We also determined the timing resolution by using the auxiliary finger PMTs as reference, and performed a data-driven study of our estimated double-peak separation capability. These studies are discussed below.

\subsection{Linearity and light yield}
\begin{figure}[hbt]
  \centering\includegraphics[width=\linewidth]{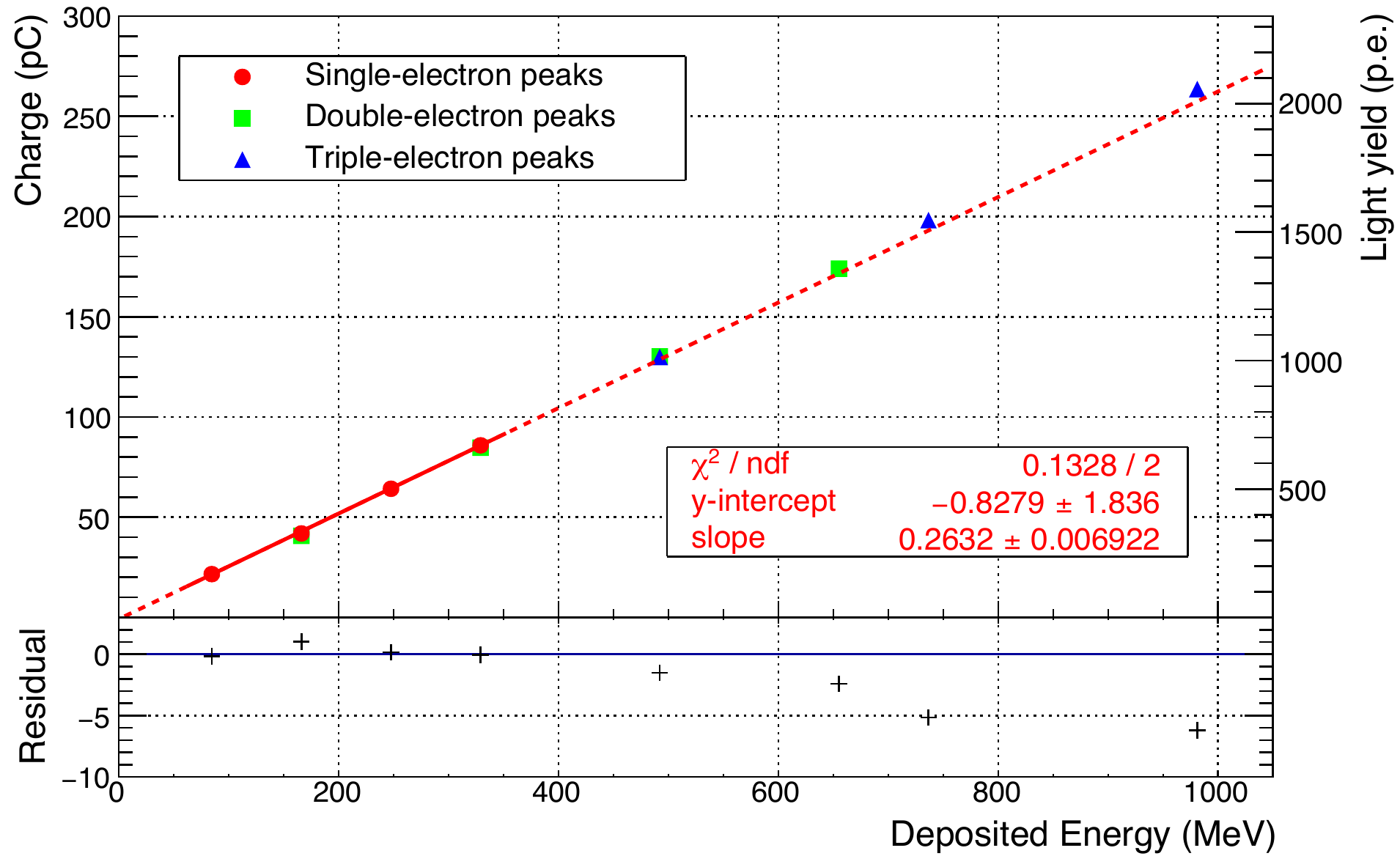}
  \caption{Detector linearity as a function of deposited energy (corrected for average leakage energy, see \cref{fig:energy_deposit}). Only single-electron peaks are fit. The residuals are also shown, demonstrating good linearity up to at least \SI{400}{\MeV}. A bias in the linearity is introduced for multiple-electron peaks due to the nature of our threshold setting and the after-pulse ringing, but does not affect the resolution since only single-electron peaks are used. The light yield is shown on the right, assuming a gain of \num{8e5} for a PMT operating voltage of \SI{1600}{\V} \cite{hpkr13478}. The obtained light yield is \SI{2.05}{\pe/\MeV}.}
  \label{fig:peak_linearity}
\end{figure}

\cref{fig:peak_linearity} shows the collected charge as a function of deposited energy. Each data point is the mean of the corresponding CB fit. The fit is performed only on single-electron peaks (labeled `Peak 1' in \cref{fig:cb_fits}), though we also plot data points corresponding to multiple-electron peaks for completeness (`Peak 2' and `Peak 3'). Since a typical single-electron pulse has a short duration and given the sampling resolution of \SI{0.2}{\ns}, only rarely do two or more electron pulses exactly overlap in time, and so multiple-electron events are not appropriate indicators of the charge linearity or energy resolution.

The plot in \cref{fig:peak_linearity} is a function of the actual deposited energy on the crystal, and not of incident beam energy. This is done to account for the fact that a single finite-sized crystal does not provide full energy containment. The correction used is shown in \cref{fig:energy_deposit}, as described in \cref{sec:sc_mc}. 

From the slope of the solid fit line in \cref{fig:peak_linearity}, we find a light yield of \SI{2.05}{\pe/\MeV}. Note that the linearity is valid only for the single-peak regime, which is what we are interested in. For double-electron peaks or higher, the ringing after-pulse is not adequately captured by our pulse-area calculation if the two peaks are close enough in time. This overestimates the total charge and introduces a bias in the linearity.

\subsection{Energy resolution}

\begin{figure}[hbt]
  \centering\includegraphics[width=\linewidth]{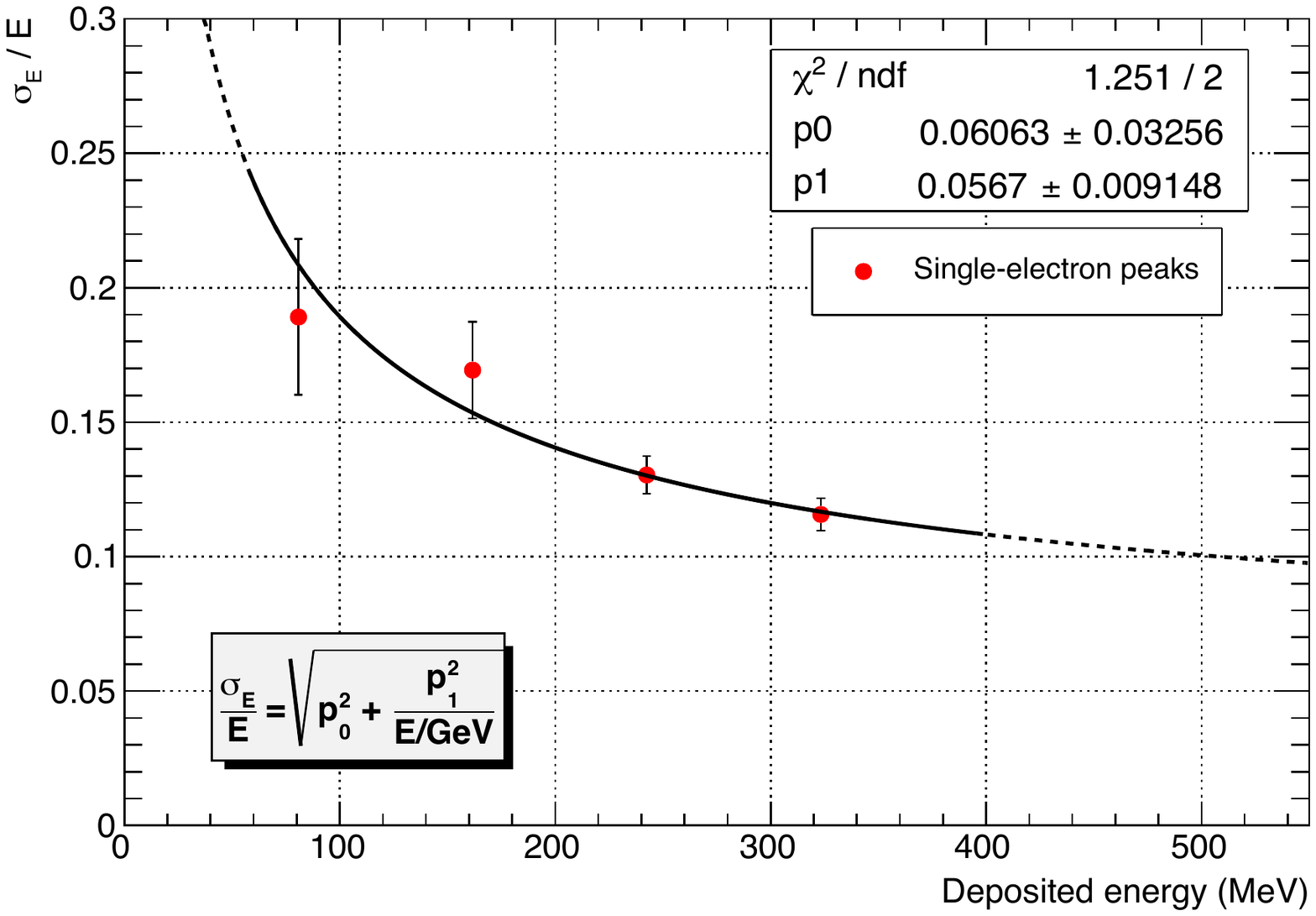}
  \caption{SAC energy resolution as a function of deposited energy. The amount of deposited energy is calculated from the incident beam energy using $f(E)$ (fraction of deposited energy) as computed with the MC simulation (\cref{fig:energy_deposit}). The fit yields a resolution of roughly 10\% at 550 MeV, which meets PADME's requirements.}
  \label{fig:energy_res}
\end{figure}

The energy resolution was calculated as the fitted sigma over mean of single-electron peaks in the charge distributions (e.g. \cref{fig:cb_fits}). \cref{fig:energy_res} shows the achieved energy resolution as a function of deposited energy on the crystal.

The data were fitted according to the expression:

\begin{equation} \frac{\sigma_E}{E} = \sqrt{p_0^2+\frac{p_1^2}{E/\mathrm{GeV}}},
\label{eq:resolution} \end{equation}


\noindent where $p_1$ and $p_2$ are two fit parameters. Note that the amount of deposited energy on the crystal (the values on the x-axis) is obtained from the incident beam energy by applying the function $f(E)$, from \cref{fig:energy_deposit}. We emphasize that this procedure can only correct for \emph{average} energy leakage, but not for \emph{fluctuations} in that quantity, which adds a significant contribution to the energy resolution.

The chosen fit function is an ad-hoc attempt to capture the overall behavior of the energy resolution curve, and not specific contributions to it. Major sources of fluctuations in the energy resolution include the aforementioned shower leakage (lateral, longitudinal, and albedo), stochastic fluctuations in the number of photo-electrons, and electronics noise. We estimate that the dominant contribution to shower leakage is the lateral lack of containment, which will be mitigated with a full calorimeter comprised of 5x5 crystals, thereby improving the energy resolution. Nevertheless, we find that with a single crystal, the energy resolution is roughly 10\% at 550 MeV, which already meets PADME's SAC requirements.

\subsection{Timing resolution}

\begin{figure}[hbt]
  \centering\includegraphics[width=\linewidth]{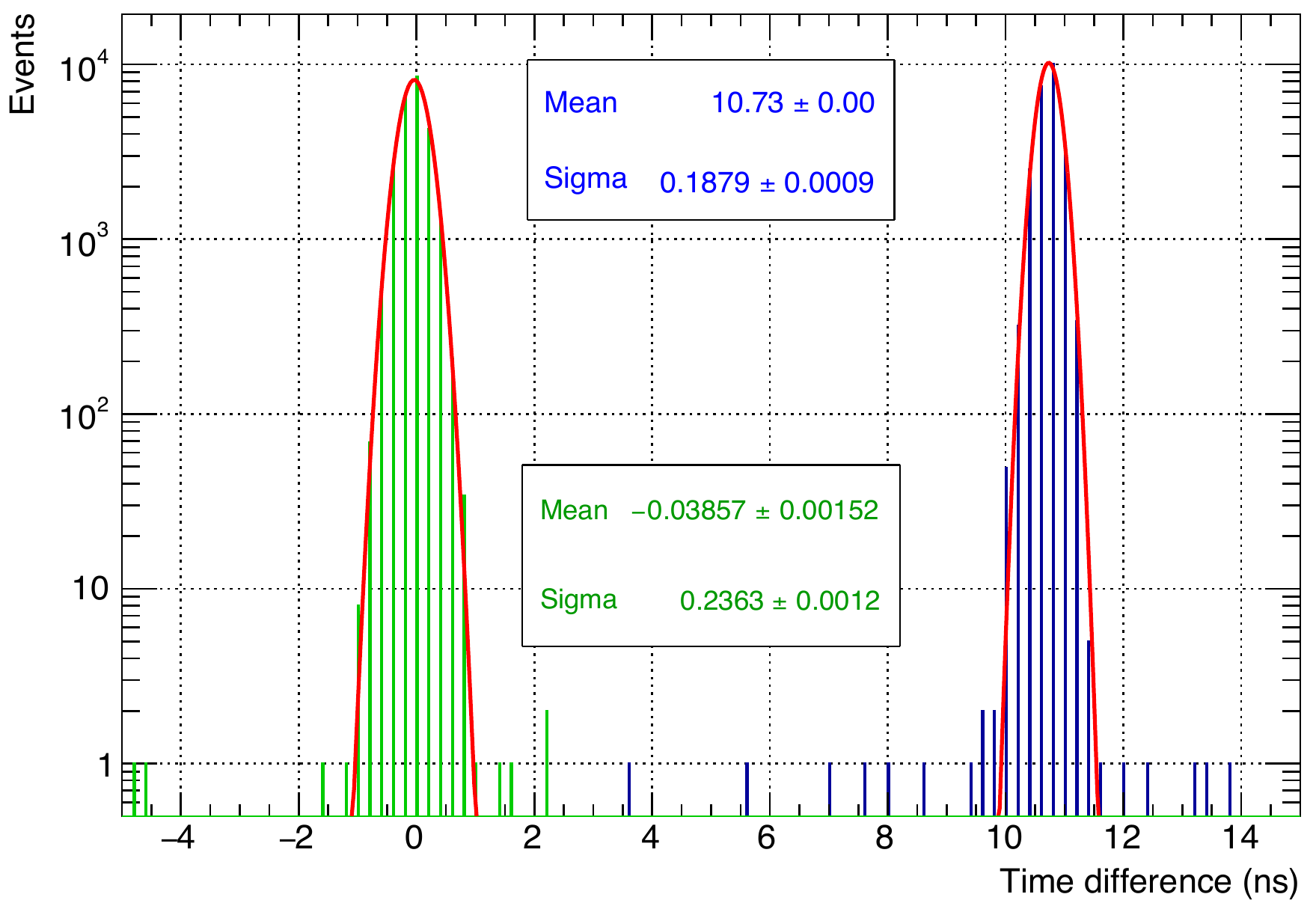}
  \caption{Difference in rise time between \pbfII{} and finger 1 channels (blue) and finger 1 and finger 2 channels (green), for a \SI{300}{\MeV} beam. The measurement is taken at 50\% of the amplitude according to a linear fit in the 20-80\% range.}
  \label{fig:time_fits}
\end{figure}

The timing resolution of the \pbfII{} + R13478UV PMT setup was determined with help from the scintillator bar and finger PMTs. First a cut was imposed on the integrated charge to select events with only one electron in them. For each selected event, the rising edge of the electron pulse was fit to a straight line, using as endpoints the 20\% and 80\% heights of the pulse amplitude. The location of the fit at 50\% height was then taken as the reference time. This procedure was done for all three channels: \pbfII{} and the two finger PMTs. The distribution of time differences between each channel was plotted. Two examples can be seen in \cref{fig:time_fits}.

\begin{figure}[hbt]
  \centering\includegraphics[width=\linewidth]{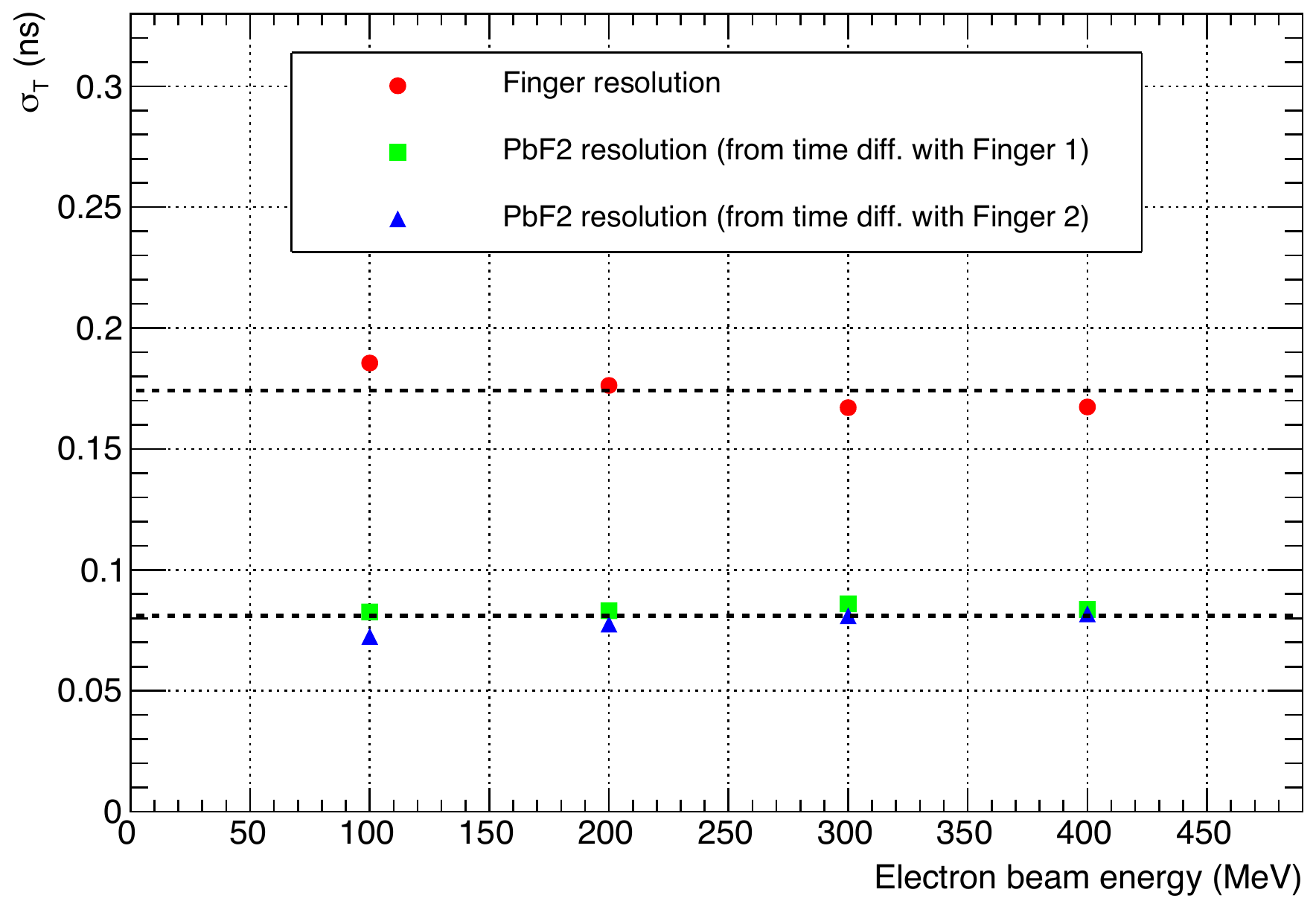}
  \caption{\pbfII{} and finger timing resolutions, calculated from the difference in rise time between the different channels. The SAC timing resolution of roughly \SI{80}{\ps} meets PADME's requirements. Note that timing resolution is not significantly affected by beam energy due to our method of estimating the arrival time. This method relies on calculating the slope of the rising pulse, whose midpoint does not vary much with amplitude.}
  \label{fig:time_res}
\end{figure}

The timing resolution was determined by extracting the sigmas of a Gaussian fit to each such distribution. From the two-finger distributions, the finger resolution was determined to be \SI{174}{\ps}. Then, from the other distributions, the \pbfII{} resolution was extracted by inserting the calculated finger resolution and summing in quadrature. The resulting resolutions for all channels and runs are shown in \cref{fig:time_res}. The \pbfII{} resolution was found to be \SI{81}{\ps}, which comfortably meets the \SI{200}{\ps} timing requirement for a successful SAC performance. 

\subsection{Double-peak separation resolution}

\begin{figure}[hbt]
  \centering\includegraphics[width=\linewidth]{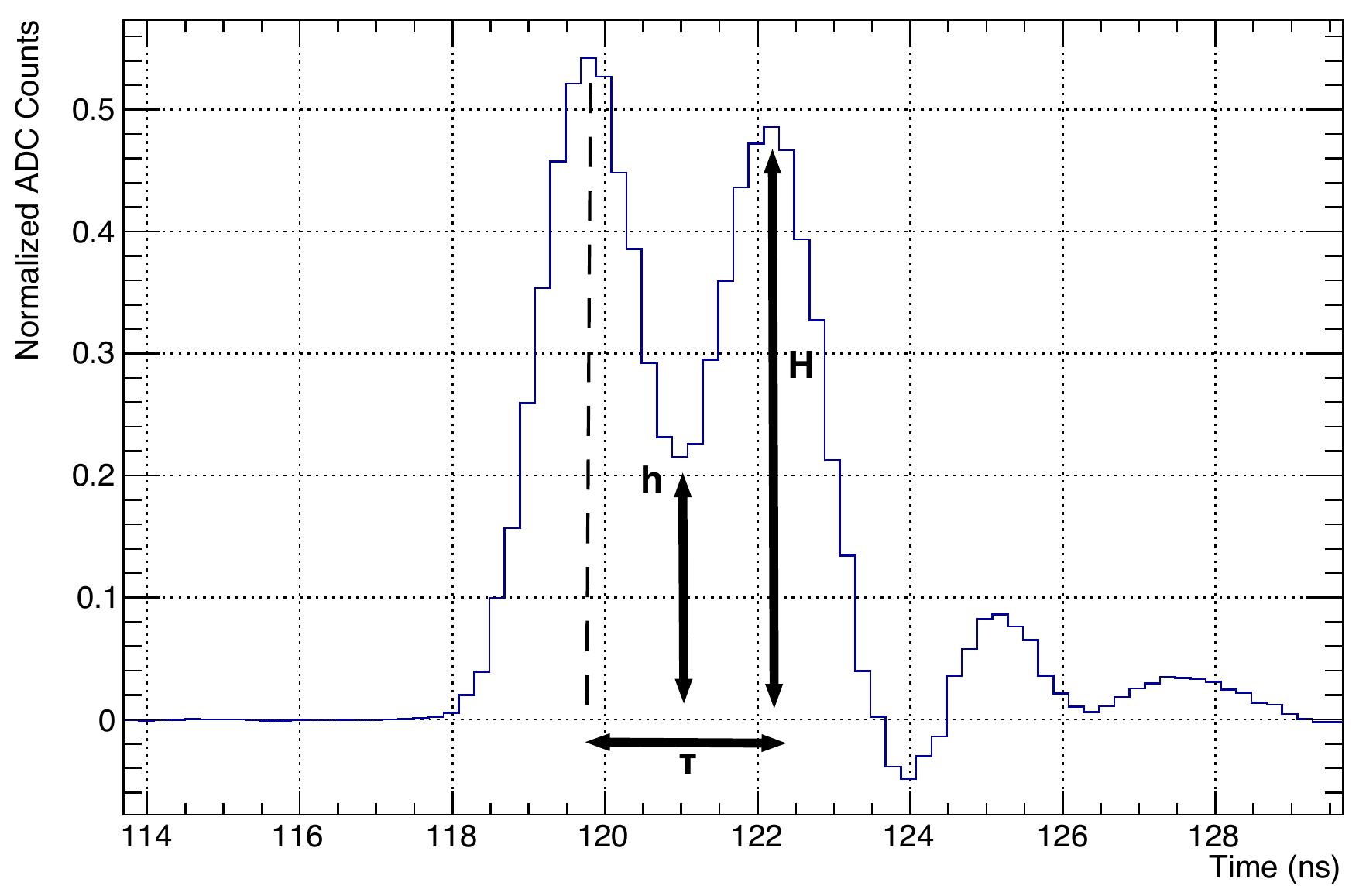}
  \caption{Sample trace used in the data-driven estimate of double-peak separation resolution. Two single-electron waveforms are overlapped with an artificially introduced separation $\tau$. The ratio $r = h/H$ between the inter-peak trough height and the second peak height characterizes the separability of the two peaks.}
  \label{fig:sample_trace_peak_separation}
\end{figure}

In addition to timing resolution, we measured the double-peak separation capability of the setup, using a purely data-driven method. From the charge distribution, a cut was imposed to select only single-electron events. Traces of selected events were then randomly added in pairs, with an artificially introduced time separation between single-electron peaks $\tau$. By varying $\tau$ we simulated increasingly overlapping pulses. For each value of $\tau$, the peak separation was measured by means of the ratio between the height of the separation trough ($h$), and the height of the smaller of two peaks ($H$). This ratio, $r=h/H$, characterizes the degree of separability between the two peaks. A value closer to 1, for example, implies the two peaks are very close together and distinguishing them is more challenging. On the other hand, a ratio closer to 0 means the peaks are far apart and identifying them is straightforward. A sample trace showing the relevant quantities defined above can be seen in \cref{fig:sample_trace_peak_separation}.

\begin{figure}[hbt]
  \centering\includegraphics[width=\linewidth]{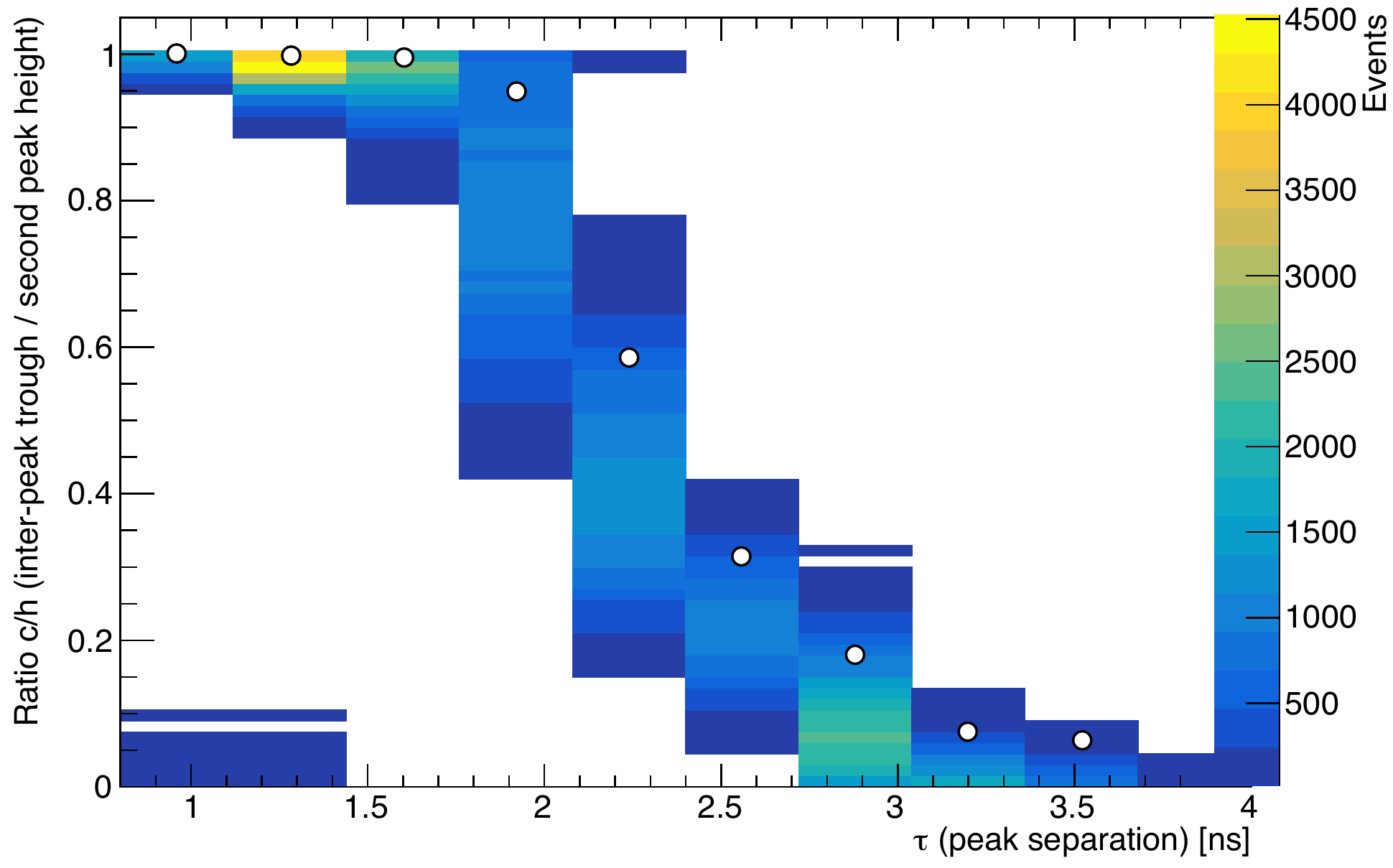}
  \caption{Ratio $r=h/H$ for different values of $\tau$. This ratio informs the separability between double-electron peaks, with a value near 1 (0) implying challenging (easier) separation. The white dots represent the 90-th percentile of each distribution.}
  \label{fig:peak_separation}
\end{figure}

\cref{fig:peak_separation} shows the distribution of ratios $r$ for a \SI{300}{\MeV} beam. The white dots correspond to the 90-th percentile of the ratio distribution for each $\tau$. Assuming a minimum separation capability of $r < 0.95$, then roughly 90\% of peaks with \SI{1.8}{\ns} separation can be distinguished. As a conservative estimate we take this to be the double-peak separation resolution. 

With a more sophisticated algorithm (for example, template fitting), it is likely that this capability can be improved further. Even when there is no local minimum (i.e. the peaks are too close together), in which case this algorithm fails, template fitting might still be able to identify a broader shoulder as a second peak. Nevertheless, this separation resolution already meets the PADME tagging requirements.

\section{Conclusions}

We have characterized the performance of a prototype of the Small-Angle Calorimeter in the PADME experiment, set to start taking data in the summer of 2018. Part of that characterization stems from a test beam with a single-crystal prototype of the SAC performed at LNF's BTF facility, in Frascati, Italy. The energy, timing, and double-peak separation resolutions all meet the demanded specifications in order to effectively tag and veto 2- and 3-gamma events from positron-electron interactions. The \pbfII{} crystal coupled to the newly developed Hamamatsu R13478UV PMT has been found to provide a light yield of \SI{2.05}{\pe/\MeV}, an energy resolution of 10\% at \SI{550}{\MeV}, and a timing resolution of \SI{81}{\ps}, with a double-peak separation capability of \SI{1.8}{\ns}. The robust double-peak resolution achieved suggests that the present setup might be suitable for a variety of high-intensity applications, being able to cope with rates higher than \SI{100}{\mega\hertz}.

We also investigated other possible detector solutions, and in particular we recommend the further consideration of a two-PMT setup using the ultra-compact Hamamatsu R9880-U110. This would allow an independent measurement of each PMT efficiency and provide a separate timing reference. This solution is anticipated for higher-energy applications, in which light yield is not a crucial factor for a satisfactory performance. 





\section*{Acknowledgements}
This work is partly supported by the Italian Ministry of Foreign Affairs and International Cooperation (MAECI), CUP I86D16000060005. GG and VK acknowledge partial support from BG-NSF DN-08-14/14.12.2016 and from LNF-SU MoU 70-06-497/07-10-2014.

The authors thank D.W. Herzog and the g-2 Collaboration for loaning the \pbfII{} crystals used in the test beam, and Robin Bjorkquist for providing some of the code used in the optical Geant4 simulation.

\input{SAC_paper.bbl}







\end{document}

%% file: SAC_paper.bbl
\providecommand{\href}[2]{#2}\begingroup\raggedright\endgroup